\documentclass[12pt]{article}
\usepackage{amsmath}
\usepackage{amssymb}
\usepackage[mathscr]{eucal}
\usepackage[usenames]{color}
\usepackage[latin5]{inputenc}
\usepackage{lmodern}
\usepackage{url}
\usepackage[
bookmarks=true,
backref=true,
colorlinks=true,
linkcolor=red,
urlcolor=green]{hyperref}

\newtheorem{thm}{Theorem}
\newtheorem{cor}{Corollary}
\newtheorem{rmk}{Remark}

\numberwithin{equation}{section}
\numberwithin{thm}{section}
\numberwithin{lemma}{section}
\numberwithin{prop}{section}
\numberwithin{cor}{section}
\numberwithin{rmk}{section}
\numberwithin{defn}{section}

\newcommand{\gen}[1]{\partial_{#1}}

\newcommand{\curl}[1]{ \left\{#1\right\} }

\newcommand{\semi}{\supset \hskip -5mm +\,}

\definecolor{darkolivegreen}{rgb}{0.333333, 0.419608, 0.1843140}

\DeclareMathOperator{\Sl}{sl} 
 
\DeclareMathOperator{\SL}{SL}

\newcommand{\dx}{\partial_x}
\newcommand{\dy}{\partial_y}

\newcommand{\dt}{\partial_t}
\newcommand{\du}{\partial_u}

\newcommand{\ds}{\partial_s}
\newcommand{\df}{\partial_{\phi}}

\begin{document}

\title{\Large Dispersionless Davey--Stewartson system: Lie symmetry algebra, symmetry group and exact solutions
}

\author{
Faruk G\"{u}ng\"{o}r\thanks{e-mail: gungorf@itu.edu.tr} \quad    Cihangir \"{O}zemir\thanks{e-mail: ozemir@itu.edu.tr}\\
\small Department of Mathematics, Faculty of Science and Letters,\\
\small Istanbul Technical University, 34469 Istanbul,
Turkey}

\date{}

\maketitle

\begin{center}
\small  In memory of Prof. Pavel Winternitz: A great mentor, collaborator and friend.
\end{center}

\begin{abstract}
Lie symmetry algebra of the dispersionless Davey--Stewartson (dDS) system is shown to be infinite-dimensional. The structure of the algebra turns out to be Kac--Moody--Virasoro one, which is typical for integrable evolution equations in $2+1$-dimensions.   Symmetry group transformations are constructed using a direct (global) approach. They are split into both connected and discrete ones. Several exact solutions are obtained as an application of the symmetry properties.

\end{abstract}

\vspace{0.5cm}

\section{Introduction}
The aim of this work is to study the \emph{semi-classical} or \emph{dispersionless} limit of the Davey-Stewartson system in (2+1) dimensions from a group-theoretical point of view. A very recent work \cite{yi2020dispersionless}, followed by \cite{yi2018dispersionless}, considers the following $\varepsilon$-dependent form of the Davey--Stewartson (DS) system parametrized by $\varepsilon$
\begin{subequations}\label{DS2}
\begin{eqnarray}
   \label{DS2a}  i\varepsilon q_t+\frac{\varepsilon^2}{2}(q_{xx}+\sigma q_{yy})+\delta q \phi &=&0, \\
   \label{DS2b}  \sigma \phi_{yy} - \phi_{xx}+(|q|^2)_{xx}+\sigma (|q|^2)_{yy} &=&0,
\end{eqnarray}
\end{subequations}
where $q(x,y,t)$ is a complex wave function and $\phi(x,y,t)$ is a real (mean-flow) function.
Eq. \eqref{DS2} with $\sigma=1$ is called the DS-I system and with $\sigma =-1$ it is called the DS-II system. The case $\delta=1$ is the focusing case and $\delta  = -1$ the defocusing case. If we replace $\phi\rightarrow \phi+ |q|^2$ in \eqref{DS2}, we obtain
\begin{subequations}\label{DS3}
\begin{eqnarray}
   \label{DS3a}  i\varepsilon q_t+\frac{\varepsilon^2}{2}\,q_{xx}+\frac{\varepsilon^2 \sigma}{2} \,q_{yy} &=& -\delta q |q|^2-\delta q \phi, \\
   \label{DS3b} \phi_{xx}-\sigma \phi_{yy} &=& 2\sigma (|q|^2)_{yy}
\end{eqnarray}
\end{subequations}
which has the form of the DS system studied in \cite{champagnewinternitz1988}
\begin{subequations}\label{DS}
\begin{eqnarray}
   \label{DS1a}  i\psi_t+\psi_{xx}+\epsilon_1\psi_{yy}&=&\epsilon_2|\psi|^2\psi+\psi w, \\
   \label{DS1b}  w_{xx}+\delta_1w_{yy}&=&\delta_2(|\psi|^2)_{yy},
\end{eqnarray}
\end{subequations}
where $\delta_1,\delta_2,\epsilon_1$ and $\epsilon_2$ are real constants, $\epsilon_1=\mp 1$, $\epsilon_2=\mp1$. This system is integrable only when $\delta_1=-\epsilon_1$.

The author of \cite{yi2020dispersionless} applies the Madelung type transformation
\begin{equation}\label{WKB}
q=\sqrt{u}\exp(i \frac{s}{\varepsilon}), \qquad u>0, \quad  s \in \mathbb{R}
\end{equation}
in \eqref{DS2}, where $u(x,y,t)>0$ is a real amplitude function and $s(x,y,t)$ is the real phase function,  and  obtains the following two different systems in the dispersionless (semiclassical) limit $\varepsilon \rightarrow 0$
\begin{subequations}\label{dDS}
\begin{eqnarray}
   \label{dDSa}  u_t + (us_x)_x+\sigma (us_y)_y&=&0, \\
   \label{dDSb} s_t+\frac{1}{2}(s_x^2+\sigma s_y^2)-\delta \phi &=&0, \\
   \label{dDSc} \sigma \phi_{yy}-\phi_{xx}+u_{xx}+\sigma u_{yy} &=&0.
 \end{eqnarray}
\end{subequations}
We take this opportunity to correct that the parameter $\sigma$ in  \eqref{dDSa} has been omitted in formula (24) of Ref. \cite{yi2020dispersionless}.
In \eqref{dDS}, $\sigma= 1$ corresponds to the dispersionless DS-I (dDS-I) system  and $\sigma = -1$ corresponds to the
dispersionless DS-II (dDS-II) system.
As has been shown in  \cite{yi2020dispersionless, yi2018dispersionless}, a remarkable fact is that the dDS system arises from the commutation condition of the Lax pair of one-degree-of-freedom Hamiltonian vector fields.  A new hierarchy (the so-called dDS hierarchy) related to dDS system has been also  defined in \cite{yi2018dispersionless}.

The semi-classical limit of the NLS equation in arbitrary space dimension is obtained in \cite{jin1994behavior, jin1999semiclassical, grenier1998semiclassical} as follows. They consider the family of Cauchy problems
\begin{subequations}\label{eSch}
\begin{eqnarray}
   \label{eScha}  &&i\hbar\Psi_t^{\hbar} +\frac{\hbar^2}{2}\Delta \Psi^{\hbar}+f(|\Psi^{\hbar}|^2)\Psi^{\hbar}=0, \qquad x\in \mathbb{R}^d, \quad t\in  \mathbb{R^+}, \\
   \label{eSchb}  &&\Psi^{\hbar}(x,0)=A(x)\exp\Big(\frac{i}{\hbar} S(x) \Big)
\end{eqnarray}
\end{subequations}
which are parametrized by $\hbar$ and supposedly have solutions $\Psi^{\hbar}(x,t)$. Investigating the semi-classical limit of this problem is to \emph{"determine the limiting behaviour of any $\Psi^{\hbar}(x,t)$ as $\hbar \rightarrow 0$" } \cite{jin1994behavior, jin1999semiclassical}. Also of both mathematical and physical interest is the existence of the limiting behaviour of the conserved densities of mass, momentum and energy. Refs. \cite{jin1994behavior, jin1999semiclassical, grenier1998semiclassical} focus on a query on validity of this limiting case  before any breakdown occurs in the solutions.

We would like to mention some of the discussions reported in \cite{klein2014numerical} regarding nonlinear Schr\"{o}dinger equation, Davey--Stewartson system and their semi-classical limit as we find them useful for  an understanding of the results available in the current manuscript. Ref. \cite{klein2014numerical} notes   the  nonlinear Schr\"{o}dinger equation
\begin{equation}\label{Sch}
i\epsilon \Psi_t+\frac{\epsilon^2}{2}\Delta \Psi-\frac{\rho}{\nu}|\Psi|^{2\nu}\Psi=0, \qquad x\in \mathbb{R}^d, \quad t\in \mathbb{R}
\end{equation}
in which $\Psi$ is a complex-valued function, $\Delta$ is the Laplace operator in $d$ dimensions, and $\epsilon \ll 1$ is a real, positive parameter. The power of nonlinearity is represented by the constant $1<2\nu <\infty$. The case $\rho=1$ is called the defocusing case and   $\rho=-1$ corresponds to the focusing case. Ref. \cite{gasser2000review} says that \emph{"It is a fundamental principle in quantum mechanics that, when the time and distance scales are large enough relative to the Planck constant $\hbar$, the system will approximately obey the laws of classical, Newtonian mechanics. This is usually rephrased in a colloquial form as: in the limit as $\hbar \rightarrow 0$ quantum mechanics becomes Newtonian mechanics. The asymptotics of quantum variables as $\hbar\rightarrow 0$ are known as 'semiclassical', which expresses this limiting behaviour."}

Since the constant $\epsilon$ in \eqref{Sch} is located at the same places with the Planck constant $\hbar$ in
linear Schr\"{o}dinger equation of quantum mechanics, the limit $\epsilon\rightarrow 0$ is considered as the semiclassical limit.  When one uses the WKB ansatz/Madelung transformation $\Psi=\sqrt{u}e^{iS/\epsilon}$ in \eqref{Sch}, the system
\begin{subequations}\label{}
\begin{eqnarray}
   \label{}  u_t+\nabla\cdot(u\nabla S)&=&0, \\
   \label{} S_t+\frac{1}{2}|\nabla S|^2+\frac{\rho}{\nu}u^\nu &=& \frac{\epsilon^2}{2}\frac{\Delta (\sqrt{u})}{\sqrt{u}}.
\end{eqnarray}
\end{subequations}
is obtained. Letting $\epsilon \rightarrow 0 $, the semi-classical limit of the NLS equation \eqref{Sch} is obtained \cite{klein2014numerical}.  Ref. \cite{klein2014numerical} considers the Davey--Stewartson system
\begin{subequations}\label{DS1}
\begin{eqnarray}
   \label{} i\epsilon \Psi_t +\epsilon^2 \Psi_{xx}-\alpha\epsilon^2 \Psi_{yy}+2 \rho(\Phi+|\Psi|^2)\Psi&=&0,\\
   \label{} \Phi_{xx}+\beta \Phi_{yy}+2(|\Psi^2|)_{xx}&=&0
\end{eqnarray}
\end{subequations}
in which $\alpha$, $\beta$ and $\rho$ assume the values $\mp 1$, and $\epsilon$ is the dispersion parameter which is small. For $\alpha=\beta=-1$ \eqref{DS1} is called DS-I equation, and for $\alpha=\beta=1$ it is called DS-II equation. DS-I equation is  integrable.  The main focus of Ref. \cite{klein2014numerical} is to study numerically the dispersionless limit of the  DS II system which appears nonlocally due to the  operator $\mathcal{D}_{+}^{-1}$
\begin{subequations}\label{dDSII}
\begin{eqnarray}
   \label{}  u_t+2 (uS_x)_x-2(uS_y)_y&=&0, \\
   \label{} S_t+S_x^2-S_y^2+2\rho \mathcal{D}_{+}^{-1}\mathcal{D}_{-}(u)&=&0
\end{eqnarray}
\end{subequations}
with $\mathcal{D}_{\pm}=\partial_x^2 \pm \partial_y^2$.

Ref. \cite{miller1998semiclassical} performs numerical experiments to study the semi-classical limit for the focusing cubic nonlinear Schr\"{o}dinger equation with analytic initial data. Ref. \cite{bronski1999numerical} also carried out numerical analyses which were in agreement with the results of \cite{miller1998semiclassical} and obtained a train of solitons.  \cite{benci2002semiclassical} considers the NLS equation in two-dimensions with a general nonlinearity including a radially symmetric potential and studies the semi-classical limit. We see further treatments in \cite{carles2007wkb}, \cite{carles2007geometric}; and we also would like to mention the book \cite{carles2020semi} by the same author on NLS equations.

The Benney system of equations, originally introduced in \cite{benney1973some}, has been obtained in \cite{zakharov1980benney, zakharov1981benney}  as  the quasiclassical limit of NLS equations, which are integrable by the inverse scattering method.
Dispersionless limits of several equations in (1+1), (2+1) and (3+1) dimensions including Heisenberg
ferromagnet equation, Landau--Lifshitz equation, KdV, NLS and DS equations are listed in \cite{myrzakulova2019}.
A discussion of dispersionless limits of integrable systems can be found in \cite{zakharov1994dispersionless},  which involves Kadomtsev--Petviashvili (KP), generalized Benney and Davey--Stewartson (DS) equations. As we mentioned, the dDS system under investigation in this manuscript is introduced in \cite{yi2020dispersionless}. As a continuation of this work,  a new
hierarchy of compatible PDEs defining infinitely many symmetries, which
is associated with the dDS system is presented in \cite{yi2018dispersionless}. Ref.  \cite{konopelchenko2007quasiclassical} presents a  generalized Weierstrass-type representation for highly corrugated
surfaces with a slow modulation in the four- and three-dimensional Euclidean spaces, which find applications in applied mathematics, string theory, membrane theory and various other fields \cite{konopelchenko2007quasiclassical}. It is stated in \cite{konopelchenko2007quasiclassical} that  integrable deformations of such corrugated surfaces in $\mathbb{R}^4$ and $\mathbb{R}^3$ are induced by the hierarchy of dispersionless DS (dDS) equations and dispersionless modified Veselov--Novikov equations. In this context, the authors obtain the quasi-classical (i.e., dispersionless) limit of the Davey--Stewartson systems.

Davey--Stewartson system \eqref{DS} was obtained in \cite{daveystewartson} in two space dimensions by a perturbation expansion  for the evolution of a wave packet a water of finite depth, $\psi$ and $w$ standing for short wave and long wave modulation amplitude functions. It is a special case of the Benney--Roskes system obtained in \cite{benneyroskes}. The generalized Davey--Stewartson (GDS) system in (2+1) dimensions  which governs the dynamics of \emph{"a  short transverse wave, a long transverse wave and a long longitudinal wave"} was derived in \cite{babaoglu2004two}. In (3+1) dimensions, we can cite  \cite{carbonaro2012three} as the derivation of a DS system in a collisionless
unmagnetized plasma via multiple-scale asymptotic expansion method for three-dimensional modulation of an electron-acoustic wave.

Lie symmetry algebra of the DS equations was shown in \cite{champagnewinternitz1988} to be  an infinite-dimensional Lie algebra isomorphic to the Kac--Moody--Virasoro (KMV) algebra exactly in the integrable case.  The variable-coefficient
DS (VCDS) system has been considered in \cite{gungor2016variable} to give a Lie symmetry analysis and in particular to examine the conditions for the VCDS system to be transformable to the standard integrable DS system. Lie algebra of the GDS system is also infinite-dimensional, as studied in \cite{gungor2006generalized, li2008symmetry}. In the (3+1)-dimensional case, Virasoro part of the Lie algebra degenerates to a finite-dimensional subalgebra and hence the full invariance algebra is an indecomposable one as semi-direct sum of this subalgebra with an infinite-dimensional ideal \cite{ozemir2020davey}. It is also our intention here to analyse  system \eqref{dDS} from a group-theoretical perspective by exploring its Lie symmetry algebra and group.

The paper is organized as follows.  In Section 2 we find the Lie symmetry algebra of the dDS system \eqref{dDS} using the infinitesimal approach. Section 3 is devoted to the derivation of   symmetry group transformations of \eqref{dDS} by the global (not infinitesimal) method. Section 4 provides reductions via a one-dimensional and two-dimensional subalgebras and their analysis to obtain exact solutions in some cases. The parameter $\sigma$ is restricted to $\sigma = \mp 1$ throughout the paper.

\section{Lie symmetry algebra}
The Lie symmetry algebra $L$ of the dDS system \eqref{dDS} will be realized in terms of vector fields of the form
\begin{equation}\label{VF-1}
V=\tau\dt+\xi\dx+\eta\dy+\zeta_1\du+\zeta_2\ds+\zeta_3\df,
\end{equation}
where $\tau$, $\xi$, $\eta$, $\zeta_1$, $\zeta_2$ and $\zeta_3$ are functions of $t,x,y,u,s,\phi$ to be determined from the infinitesimal invariance condition.
Applying this condition and solving an overdetermined system of linear PDEs we find that a general element of $L$ can be written in the form
\begin{equation}\label{VF}
V=\mathscr{X}(\tau)+\mathscr{Y}(g)+\mathscr{Z}(h)+\mathscr{W}(m)+\mu_0\mathscr{D}_0,
\end{equation}
where $\mu_0$ is an arbitrary constant and
\begin{subequations}\label{gen}
\begin{eqnarray}
\mathscr{D}_0 &=& x\dx+y\dy+2u\du+2s\ds+2\phi\df,\\
\mathscr{X}(\tau)&=&\tau(t)\dt+\frac{\dot\tau(t)}{2}\big(x\dx+y\dy\big)-\dot\tau(t) u \du+\frac{\ddot\tau(t)}{4}\big(x^2+\sigma y^2\big)\ds  \nonumber \\
       &+&\big[\frac{\dddot\tau(t)}{4\delta}\big(x^2+\sigma y^2\big)-\dot{\tau}(t)\phi\big]\df,\\
\mathscr{Y}(g)&=& g(t)\dx+x\dot g(t)\ds+\frac{x}{\delta} \ddot{g}(t)\df,\\
\mathscr{Z}(h)&=& h(t)\dy+\sigma y \dot h(t)\ds+\frac{y}{\sigma \delta}\ddot h(t)\df,\\
\mathscr{W}(m)&=& m(t)\ds+\frac{1}{\delta} \, \dot m(t)\df \, .
 \end{eqnarray}
\end{subequations}
The functions $\tau(t)$, $g(t)$, $h(t)$ and $m(t)$ figuring in \eqref{VF} are arbitrary.

The commutation relations between them are given as follows
\begin{equation}
[\mathscr{Y}(g),\mathscr{D}_0]=\mathscr{Y}(g),\quad [\mathscr{Z}(h),\mathscr{D}_0]=\mathscr{Z}(h), \quad [\mathscr{W}(m),\mathscr{D}_0]=2\mathscr{W}(m)
\end{equation}
and
\begin{subequations}
\begin{eqnarray}
&&[\mathscr{X}(\tau_1),\mathscr{X}(\tau_2)]=\mathscr{X}(\tau_1\dot\tau_2-\tau_2\dot\tau_1),\\
&&[\mathscr{Y}(g_1),\mathscr{Y}(g_2)]=\mathscr{W}(g_1 \dot g_2-g_2\dot g_1),\\
&&[\mathscr{Z}(h_1),\mathscr{Z}(h_2)]=\mathscr{W}\big(\sigma(h_1 \dot h_2-h_2 \dot h_1)\big),\\
&&[\mathscr{X}(\tau),\mathscr{Y}(g)]=\mathscr{Y}(\tau  \dot g-\frac{1}{2}g\dot\tau),   \label{XY}\\
&&[\mathscr{X}(\tau),\mathscr{Z}(h)]=\mathscr{Z}(\tau \dot h-\frac{1}{2} h \dot\tau),\\
&&[\mathscr{X}(\tau),\mathscr{W}(m)]=\mathscr{W}(\tau \dot m)     \label{XW}
\end{eqnarray}
\end{subequations}
and the vanishing ones are, just to note,
\begin{equation}
[\mathscr{D}_0,\mathscr{X}(\tau)]=[\mathscr{W}(m_1),\mathscr{W}(m_2)]=[\mathscr{Y}(g),\mathscr{Z}(h)]=[\mathscr{Y}(g),\mathscr{W}(m)]=[\mathscr{Z}(h),\mathscr{W}(m)]=0.
\end{equation}

\begin{thm}
System \eqref{dDS}  admits an infinite dimensional Lie symmetry algebra $L$ with the infinitesimal generator \eqref{VF} depending on four arbitrary smooth functions $\tau(t)$, $g(t)$, $h(t)$ and $m(t)$. The algebra has the Levi decomposition of a simple algebra and its radical
    \begin{equation}
    L=\{\mathscr{X}(\tau)\}\semi\{\mathscr{Y}(g),\mathscr{Z}(h),\mathscr{W}(m),\mathscr{D}_0\}.
    \end{equation}
        The algebra  $\{\mathscr{X}(\tau)\}\semi\{\mathscr{Y}(g),\mathscr{Z}(h),\mathscr{W}(m)\}$ is isomorphic to the DS symmetry algebra of \eqref{DS},   which has a Kac--Moody--Virasoro (KMV) structure \cite{champagnewinternitz1988}.
\end{thm}

\begin{rmk}
  The presence of KMV loop structure  is typical of many integrable PDEs  in  2+1-dimensions like KP equation and its dispersionless variant and integrable DS system.
\end{rmk}

\section{Symmetry  group transformations}
In this Section we will find the symmetry  transformation  that maps \eqref{dDS} to itself.  One parameter subgroups of the full symmetry group can be obtained by integrating the vector fields in the symmetry algebra. When the symmetry algebra is infinite-dimensional then the integration  procedure can be quite tricky. Here we shall pursue an alternative direct approach to find the full symmetry group.  An added bonus is that all possible discrete symmetry transformations, not possible in infinitesimal approach, may be caught.

Inspection of the symmetry algebra \eqref{VF-1}-\eqref{gen} suggests that we may restrict the point symmetry transformations to the  fiber-preserving ones of the form
\begin{subequations}\label{transf}
\begin{eqnarray}
      u(x,y,t)=A_1(x,y,t)\,\tilde{u}(\tilde{x},\tilde{y},\tilde{t})+B_1(x,y,t),\\
      s(x,y,t)=A_2(x,y,t)\,\tilde{s}(\tilde{x},\tilde{y},\tilde{t})+B_2(x,y,t),\\
      \phi(x,y,t)=A_3(x,y,t)\,\tilde{\phi}(\tilde{x},\tilde{y},\tilde{t})+B_3(x,y,t),
 \end{eqnarray}
\end{subequations}
where $\tilde{x}=X(x,y,t)$, $\tilde{y}=Y(x,y,t)$, $\tilde{t}=T(t)$, under the invertibility condition
\begin{equation}\label{invert}
  A_1A_2A_3\neq 0, \qquad \frac{\partial (\tilde x, \tilde y)}{\partial (x,y)}\, \dot T \neq 0.
\end{equation}

We determine
the functions $X$, $Y$, $T$, $A_i$, $B_i$, $i=1,2,3$  such that  \eqref{dDS} is invariant under this transformation; that is, the transformation \eqref{transf} maps the system  \eqref{dDS} to the one in the same form with new variables replaced by tildes
\begin{subequations}\label{barDS}
\begin{eqnarray}
   \label{bara}\tilde{u}_{\tilde{t}} + (\tilde{u}\tilde{s}_{\tilde{x}})_{\tilde{x}}+\sigma(\tilde{u}\tilde{s}_{\tilde{y}})_{\tilde{y}}&=&0, \\
   \label{bars}\tilde{s}_{\tilde{t}}+\frac{1}{2}(\tilde{s}_{\tilde{x}}^2+\sigma \tilde{s}_{\tilde{y}}^2)-\delta \tilde{\phi} &=&0, \\
   \label{barf}\sigma \tilde{\phi}_{\tilde{y}\tilde{y}}-\tilde{\phi}_{\tilde{x}\tilde{x}}+\tilde{u}_{\tilde{x}\tilde{x}}+\sigma \tilde{u}_{\tilde{y}\tilde{y}} &=&0.
 \end{eqnarray}
\end{subequations}
When we plug the transformation \eqref{transf} in $\eqref{dDS}$, we obtain the transformed system \eqref{barDS}, where each equation includes several extra partial derivatives of the dependent variables.

In order that \eqref{barf} does not include the terms $\tilde{u}_{\tilde{x}\tilde{y}}$ and $\tilde{\phi}_{\tilde{x}\tilde{y}}$ and \eqref{bara}
does not include the terms $\tilde u \tilde{s}_{\tilde{x} \tilde{y}}$ and $\tilde{s}_{\tilde{y} \tilde{y}}$, one must have, respectively,
\begin{subequations}
\begin{eqnarray}
      A_1(X_xY_x+\sigma X_y Y_y)&=&0,\\
      A_3(X_xY_x-\sigma X_y Y_y)&=&0, \\
      A_1A_2(X_xY_x+\sigma X_y Y_y)&=&0,\\
      A_2 B_1 (Y_x^2+\sigma Y_y^2)&=&0.
 \end{eqnarray}
\end{subequations}
As $A_1A_2A_3\neq 0$, these conditions give two possibilities: either we have $Y_x=X_y=0$ or $X_x=Y_y=0$. In both cases we have $B_1(x,y,t)=0.$

\par{a) $X=X(x,t)$, $Y=Y(y,t)$:}

In this case, in Eq. \eqref{bars} there is a term  that does not include any of the dependent variables, which must vanish:
\begin{equation}
B_{2,t}+\frac{1}{2}B_{2,x}^2+\frac{\sigma}{2}B_{2,y}^2-\delta B_3=0
\end{equation}
so we have
\begin{equation}
B_3(x,y,t)=\frac{1}{\delta}\Big(B_{2,t}+\frac{1}{2}B_{2,x}^2+\frac{\sigma}{2}B_{2,y}^2\Big).
\end{equation}
Moreover, eliminating  the terms $\tilde s$, $\tilde s \tilde{s}_{\tilde x} $ and $\tilde s \tilde{s}_{\tilde y} $ that appear in \eqref{bars}
results in
\begin{subequations}
\begin{eqnarray}
   A_{2,t}+A_{2,x}B_{2,x}+\sigma A_{2,y}B_{2,y}&=&0,\\
   A_2 A_{2,x} X_x&=&0,\\
  A_2 A_{2,y} Y_y&=&0,
 \end{eqnarray}
\end{subequations}
therefore $A_2$ is a constant, say $A_2(x,y,t)=1/ \mu.$ After this, \eqref{bars} takes the form
\begin{equation}\label{den2}
\tilde{s}_{\tilde{t}}+\frac{X_x^2}{2\mu \dot T}\tilde{s}_{\tilde{x}}^2+\frac{\sigma Y_y^2}{2\mu \dot T} \tilde{s}_{\tilde{y}}^2
-\frac{\delta \mu A_3}{\dot T} \tilde{\phi} + \frac{1}{\dot T}(X_t+X_x B_{2,x})\tilde{s}_{\tilde{x}}+\frac{1}{\dot T}(Y_t+\sigma Y_y B_{2,y})\tilde{s}_{\tilde{y}}=0.
\end{equation}
If we compare \eqref{den2} with \eqref{bars}, we arrive at
\begin{equation}
X(x,t)=\epsilon_1\sqrt{\mu\dot{T}}x+\alpha(t), \quad Y(y,t)=\epsilon_2\sqrt{\mu \dot{T}}y+\beta(t), \quad A_3(x,y,t)=\frac{1}{\mu}\,\dot{T}(t)
\end{equation}
with $\epsilon_1^2=1$ and $\epsilon_2^2=1$. Furthermore, from the system
$$X_t+X_x B_{2,x}=0,  \quad Y_t+\sigma Y_y B_{2,y}=0$$ we find that
\begin{equation}
B_2(x,y,t)=-\frac{1}{4}\frac{\ddot{T}}{\dot T}(x^2+\sigma y^2)
-\frac{\epsilon_1 \dot \alpha}{\sqrt{\mu\dot T}}\,x-\frac{\epsilon_2 \sigma \dot \beta}{\sqrt{\mu\dot T}}\,y+\gamma(t).
\end{equation}
This achieves transformation of \eqref{dDSb} to \eqref{bars}. Subsequently, the transformed form of \eqref{dDSc} becomes
\begin{equation}
\frac{\sigma \dot T}{\mu A_1} \, \tilde{\phi}_{\tilde{y}\tilde{y}}-\frac{\dot T}{\mu A_1}\, \tilde{\phi}_{\tilde{x}\tilde{x}}+\tilde{u}_{\tilde{x}\tilde{x}}+\sigma \tilde{u}_{\tilde{y}\tilde{y}} +\frac{2\epsilon_1 A_{1,x}}{A_1 \sqrt{\mu\dot T}}\, \tilde{u}_{\tilde{x}}+\frac{2\epsilon_2 \sigma A_{1,y}}{A_1 \sqrt{\mu \dot T}}\, \tilde{u}_{\tilde{y}}+\frac{A_{1,xx}+\sigma A_{1,yy}}{\mu A_1 \dot T} \, \tilde u=0.   \,
\end{equation}
We see that we ensure the invariance of \eqref{dDSc}  with the choice $A_1(x,y,t)=\dot T /\mu$.

Summing up we obtain:

\begin{thm}
The full symmetry group $G$ of \eqref{dDS} is composed of the transformations
\begin{equation}\label{sym-a-1}
\begin{split}
   & X=\epsilon_1 \sqrt{\mu \dot T} x+\alpha(t), \quad Y=\epsilon_2 \sqrt{\mu \dot T} y+\beta(t), \quad T=T(t), \quad  \dot T>0,  \quad \mu>0,\\
    &  u=\frac{1}{\mu}\,\dot{T} \, \tilde{u}(X,Y,T),\\
 & s=\frac{1}{\mu}\,  \tilde s(X,Y,T) -\frac{1}{\sqrt{\mu\dot T}} (\epsilon_1 \dot{\alpha} x
    +\epsilon_2  \sigma \dot{\beta} y)-\frac{\ddot{T}}{4\dot T }\, (x^2 +\sigma y^2)+\gamma(t),\\
&  \phi=\frac{1}{\mu}\,\dot{T} \,\tilde{\phi}(X,Y,T)
       +\frac{\epsilon_1}{\delta\sqrt{\mu\dot T}} \Big(\dot{\alpha}\frac{\ddot{T}}{\dot{T}}-\ddot{\alpha}\Big)x
       +\frac{\epsilon_2 \sigma}{\delta\sqrt{\mu\dot T}} \Big(\dot{\beta}\frac{\ddot{T}}{\dot{T}}-\ddot{\beta}\Big)y \\
&       \qquad -\frac{1}{4\delta}\, S\{T,t\} \, (x^2+\sigma y^2)
        +\frac{1}{2\delta \mu\dot T} (\dot{\alpha}^2+\sigma \dot{\beta}^2)+\frac{\dot{\gamma}}{\delta},
\end{split}
\end{equation}
where
$T$, $\alpha, \beta, \gamma$ are arbitrary smooth functions of $t$ and $\epsilon_1^2=1$,  $\epsilon_2^2=1$. In \eqref{sym-a-1} we have defined the Schwarzian derivative of $T(t)$ with respect to $t$ as
\begin{equation}\label{schwarz}
 S\{T,t\}=\frac{\dddot{T}}{\dot T}-\frac{3}{2}\left(\frac{\ddot{T}}{\dot{T}}\right)^2.
\end{equation}

\end{thm}

\begin{cor}
When  $\epsilon_1=\epsilon_2=1$ in \eqref{sym-a-1},  we obtain the point Lie group transformation $G_0$ involving four arbitrary functions $T(t)$, $\alpha(t)$, $\beta(t)$ and $\gamma(t)$ as the connected component of the full symmetry group $G$.
The discrete component $G_D$ of $G$ is composed of the sign-changing discrete transformations
\begin{align*}
&(x,y,t,u,s,\phi)\mapsto (-x,y,t,u,s,\phi),\\
&(x,y,t,u,s,\phi)\mapsto (x,-y,t,u,s,\phi),\\
&(x,y,t,u,s,\phi)\mapsto (-x,-y,t,u,s,\phi).
\end{align*}
The full symmetry group $G$  can be expressed as the semi-direct product
\begin{equation}\label{full-sym-group}
  G=G_D\rtimes G_0,
\end{equation}
where $G_D\simeq\mathbb{Z}_2\times \mathbb{Z}_2$ is the group of discrete transformations. The group structure \eqref{full-sym-group} means that $G_0$ is an invariant subgroup of $G$.
\end{cor}

The continuous symmetry  group $G_0$  can be written in the infinitesimal form
\begin{equation}\label{infi-Q}
  (u,s,\phi)=(\tilde{u},\tilde{s},\tilde{\phi})+\varepsilon (Q_1[\tilde{u}],Q_2[\tilde{s}],Q_3[\tilde{\phi}])+\mathcal{O}(\varepsilon^2)
\end{equation}
by introducing a group parameter $\varepsilon$  into $\mu$ and  arbitrary functions $T, \alpha, \beta, \gamma$ through the relations
\begin{equation}\label{group-param}
\begin{split}
   & T(t)=t+\varepsilon \tau(t)+\mathcal{O}(\varepsilon^2),  \quad \mu=1+2\mu_0 \varepsilon+\mathcal{O}(\varepsilon^2), \\
    & \alpha(t)=\varepsilon g(t)+\mathcal{O}(\varepsilon^2),  \quad \beta(t)=\varepsilon h(t)+\mathcal{O}(\varepsilon^2), \quad \gamma(t)=\varepsilon m(t)+\mathcal{O}(\varepsilon^2)
\end{split}
\end{equation}
and expanding \eqref{sym-a-1} into Taylor series near $\varepsilon=0$,  keeping only terms linear in $\varepsilon$. The functions $Q_1$, $Q_2$, $Q_3$ will appear precisely the same as the characteristic functions defined by
$$Q_{j}[u]=Q_{j}(x,y,t,u,u_x,u_y,u_t)=\tau u_t+\xi u_x+\eta u_y-\zeta_{j},   \quad j=1,2,3$$
of the corresponding infinitesimal symmetries given in \eqref{gen}.

We can readily obtain the $\SL(2,\mathbb{R})$ symmetry group of the subalgebra
$$\Sl(2,\mathbb{R})\simeq\curl{\mathscr{X}(1), \mathscr{X}(t), \mathscr{X}(t^2)}$$  by restricting the parameters and arbitrary functions in \eqref{sym-a-1} with $(\epsilon_1,\epsilon_2)=(1,1)$ as
\begin{equation}\label{sub-case}
  \mu=1,  \quad \alpha(t)=\beta(t)=\gamma(t)=0, \quad T(t)=\frac{at+b}{ct+d}, \quad ad-bc=1.
\end{equation}

The condition $\ddot T\neq 0$ ($c\ne 0$) is true   so that $t$ transforms via  M\"obius transformations. The associated symmetry group transformations in this case depend on three group parameters $a,b,c$
\begin{equation}\label{sym-sl2}
\begin{split}
   &  X=\frac{x}{ct+d},\quad   Y=\frac{y}{ct+d}, \quad   T=\frac{at+b}{ct+d},\\
 & u=\frac{1}{(ct+d)^{2}} \, \tilde{u}(X,Y,T),\\
 & s=\tilde s(X,Y,T)    + \frac{c}{2(ct+d) }\, (x^2 +\sigma y^2),\\
&  \phi=\frac{1}{(ct+d)^{2}}  \,\tilde{\phi}(X,Y,T).
\end{split}
\end{equation}

On the other hand, if we restrict $T(t)$ to the solution of the Schwarzian equation
\begin{equation}\label{schw-eq}
  S\curl{T,t}=2\rho(t)
\end{equation}
for some $\rho$, then $T$ can be expressed as the three-parameter ratio of the linear combination of two linearly independent solutions   $\lambda(t), \nu(t)$  of the second order linear ordinary differential equation
$$\ddot{\Omega}+\rho(t)\Omega=0.$$
In this case, letting $\alpha,\beta,\gamma$ remain free, the symmetry group is somewhat more general
\begin{equation}\label{sym-gen}
\begin{split}
   &  X=\lambda_2^{-1}x+\alpha,\quad   Y=\lambda_2^{-1}y+\beta, \quad   T=\frac{\lambda_1}{\lambda_2}=\frac{a \lambda +b  \nu}{c \lambda +d  \nu}, \quad (ad-bc)W(\lambda,\nu)=-1,\\
 & u=\lambda_2^{-2} \, \tilde{u}(X,Y,T),\\
 & s=\tilde s(X,Y,T) + \frac{\dot{\lambda}_2}{2\lambda_2}\, (x^2 +\sigma y^2) -\lambda_2(\dot{\alpha}x+\sigma\dot{\beta}y)+\gamma   ,\\
&  \phi=\lambda_2^{-2} \,\tilde{\phi}(X,Y,T)-\frac{\rho}{2\delta}(x^2+\sigma y^2)-\frac{1}{\delta } [(2\dot{\alpha }\dot{\lambda}_2+\lambda_2\ddot{\alpha})x+\sigma(2\dot{\beta }\dot{\lambda}_2+\lambda_2\ddot{\beta})y] +\\
& \qquad +\frac{\lambda_2^2}{2\delta}(\dot{\alpha}^2+\sigma \dot{\beta}^2)+\frac{\dot{\gamma}}{\delta}.
\end{split}
\end{equation}
We note that we have used the fact that $W(\lambda_1,\lambda_2)=-1$ and $\dot{T}=\lambda_2^{-2}$.

Transformation formula \eqref{sym-sl2} or \eqref{sym-gen} states that if the functions $(\tilde{u},\tilde{s},\tilde{\phi})$ solve the original system  then so do  $(u,s,\phi)$. A class of seed solutions can be chosen in the form
\begin{equation}\label{seed-sol}
  \tilde{u}=U(x,y),  \quad \tilde{s}=F(t), \quad \tilde{\phi}=\delta^{-1}\dot{F}(t),
\end{equation}
where $U$ is a solution to the Laplace equation $\Delta U=0$ in the plane and $F$ is an arbitrary function. Applying this solution to transformation  \eqref{sym-gen} generates solutions in which all independent variables  are involved.

\par{b) $X=X(y,t)$, $Y=Y(x,t)$:}

Similar to the previous case, setting equal to zero a term added in \eqref{bars}  we have
\begin{equation}
B_3(x,y,t)=\frac{1}{\delta}\Big(B_{2,t}+\frac{1}{2}B_{2,x}^2+\frac{\sigma}{2}B_{2,y}^2\Big).
\end{equation}
The terms $\tilde s$, $\tilde s \tilde{s}_{\tilde x} $ and $\tilde s \tilde{s}_{\tilde y} $ that appear in \eqref{bars} are eliminated
if we put the constraints
\begin{subequations}
\begin{eqnarray}
   A_{2,t}+A_{2,x}B_{2,x}+\sigma A_{2,y}B_{2,y}&=&0,\\
   A_2 A_{2,y} X_y&=&0,\\
  A_2 A_{2,x} Y_x&=&0,
 \end{eqnarray}
\end{subequations}
from which it follows that $A_2$ should be a constant, say $A_2(x,y,t)=1/\mu.$ With this information, \eqref{bars} takes the form
\begin{equation}\label{den22}
\tilde{s}_{\tilde{t}}+\frac{\sigma X_y^2}{2\mu\dot T}\tilde{s}_{\tilde{x}}^2+\frac{Y_x^2}{2\mu \dot T} \tilde{s}_{\tilde{y}}^2
-\frac{\delta \mu A_3}{\dot T} \tilde{\phi} + \frac{1}{\dot T}(X_t+\sigma X_y B_{2,y})\tilde{s}_{\tilde{x}}+\frac{1}{\dot T}(Y_t+ Y_x B_{2,x})\tilde{s}_{\tilde{y}}=0.
\end{equation}
We find
\begin{equation}
X(y,t)=\epsilon_1\sqrt{\sigma \mu\dot{T}}y+\alpha(t), \quad Y(x,t)=\epsilon_2\sqrt{\sigma \mu  \dot{T}}x+\beta(t), \quad A_3(x,y,t)=\frac{1}{\mu}\,\dot{T}(t),
\end{equation}
where $\epsilon_1^2=1$ and $\epsilon_2^2=1$.
Furthermore, solving the pair of PDEs for $B_2$
$$X_t+\sigma X_y B_{2,y}=0, \quad Y_t+ Y_x B_{2,x}=0$$
we find
\begin{equation}
B_2(x,y,t)=-\frac{1}{4}\frac{\ddot{T}}{\dot T}(x^2+\sigma y^2)
-\frac{\epsilon_1 \sigma\dot \alpha}{\sqrt{\sigma\mu \dot T}}\, y-\frac{\epsilon_2  \dot \beta}{\sqrt{\sigma\mu\dot T}} \, x+\gamma(t)
\end{equation}
and so the system \eqref{dDSb} is mapped to \eqref{bars}. Now let us have a look at how \eqref{dDSc} has transformed:
\begin{equation}
-\frac{\sigma \dot T}{\mu A_1} \tilde{\phi}_{\tilde{y}\tilde{y}}+\frac{\dot T}{\mu A_1}\tilde{\phi}_{\tilde{x}\tilde{x}}+\tilde{u}_{\tilde{x}\tilde{x}}+\sigma \tilde{u}_{\tilde{y}\tilde{y}} +\frac{2\epsilon_1 \sigma \sqrt{\sigma \mu \dot T} A_{1,y}}{\mu A_1 \dot T}\tilde{u}_{\tilde{x}}+\frac{2\epsilon_2 \sqrt{\sigma \mu\dot T}A_{1,x}}{\mu A_1 \dot T}\tilde{u}_{\tilde{y}}+\frac{A_{1,xx}+\sigma A_{1,yy}}{\mu A_1 \dot T} \tilde u=0.
\end{equation}
For invariance of \eqref{dDSb} it is sufficient to choose $A_1(x,y,t)=-\dot{T}(t)/\mu$. Then, the invariance of \eqref{dDSa} is automatically satisfied without any further condition.

Finally, we find that the associated symmetry transformation involves four arbitrary functions $T(t)$, $\alpha(t)$, $\beta(t)$ and $\gamma(t)$ having the form
\begin{equation}\label{sym-b}
\begin{split}
   & X=\epsilon_1\sqrt{\sigma\mu\dot{T}}y+\alpha(t), \quad Y=\epsilon_2\sqrt{\sigma \mu  \dot{T}}x+\beta(t), \quad T=T(t), \\
 &  u=-\frac{1}{\mu}\,\dot{T} \, \tilde{u}(X,Y,T),\\
 &  s= \frac{1}{\mu}\,\tilde s(X,Y,T) -\frac{1}{\sqrt{\sigma\mu \dot T}} (\epsilon_1\sigma  \dot{\alpha} y
    +\epsilon_2  \dot{\beta} x)-\frac{\ddot{T}}{4\dot T }\, (x^2 +\sigma y^2)+\gamma(t),\\
 & \phi=\frac{1}{\mu}\,\dot{T} \,\tilde{\phi}(X,Y,T)
       +\frac{\epsilon_1}{\delta\mu}\sqrt{\sigma \mu \dot T}\Big(\frac{\dot{\alpha}\ddot{T}}{\dot{T}^{2}}-\frac{\ddot{\alpha}}{\dot{T}}\Big)y
       +\frac{\epsilon_2 \sigma }{\delta\mu}\sqrt{\sigma \mu \dot T}\Big(\frac{\dot{\beta}\ddot{T}}{\dot{T}^{2}}-\frac{\ddot{\beta}}{\dot{T}}\Big)x \\
       &-\frac{1}{4\delta}\, S\{T,t\} \, (x^2+ \sigma y^2)
        +\frac{1}{2\delta\mu \dot T} (\dot{\alpha}^2+ \sigma\dot{\beta}^2)+\frac{1}{\delta} \dot{\gamma}.
\end{split}
\end{equation}
We note that transformations \eqref{sym-b} will only induce discrete ones as they cannot be continuously connected to the identity transformation. For instance, system \eqref{dDS} is invariant under the discrete group of $x\leftrightarrow y$ coordinate exchanging and dependent variable sign-changing transformation
$$(x,y,t,u,s,\phi)\mapsto (y,x,t,-u,s,\phi).$$

\section{Reductions and exact solutions}
In this Section we construct some exact solutions invariant under the one-dimensional subalgebra $\mathscr{Y}(g)+\mathscr{W}(m)$ and also two-dimensional subalgebras obtained by embedding $\mathscr{X}(1)$ and $\mathscr{X}(t)$ into $\mathscr{Y}(g)+\mathscr{W}(m)$.

\subsection{Reduction by the subalgebra $\mathscr{Y}(g)+\mathscr{W}(m)$}
First we perform reduction by the subalgebra $\widehat{V}=\mathscr{Y}(g)+\mathscr{W}(m)$, $g\ne 0$.
We observe that $g(t)$ can be removed from $\widehat{V}$ by an application of the symmetry group generated by $\exp\curl{\varepsilon \mathscr{Y}(G)}$
\begin{equation}\label{sym-group-Y}
  \begin{split}
      & \tilde{x}=x+\varepsilon G(t),  \quad \tilde{y}=y, \quad \tilde{t}=t,  \quad \tilde{u}=u, \\
       & \tilde{s}=s+\varepsilon\left(x+\frac{\varepsilon}{2}\right)\dot{G}(t), \quad \tilde{\phi}=\phi+\frac{\varepsilon}{\delta}\left(x+\frac{\varepsilon}{2}\right)\ddot{G}(t)
   \end{split}
\end{equation}
with $G$ defined by
$$G(t)=\frac{g}{\varepsilon}\int_{t_0}^t (g^{-1}m)(z)dz+c g(t),$$ where $c$ is a constant.
Thus $\widehat{V}$ is  conjugate to the gauge subgroup $\mathscr{W}(m)$. But this does not lead to any reduction so we prefer to keep $g(t)$ as it is.

Solutions of \eqref{dDS} invariant under the group of transformations generated by  $\mathscr{Y}(g)+\mathscr{W}(m)$ will have the form
\begin{subequations}
\begin{eqnarray}
       u&=&U(y,t),\\
       s&=&S(y,t)+\frac{\dot g(t)}{2g(t)}x^2+\frac{m(t)}{g(t)}x, \\
    \phi&=&\Phi(y,t)+\frac{\ddot g(t)}{2\delta g(t)}x^2+\frac{\dot m(t)}{\delta g(t)}x .
 \end{eqnarray}
\end{subequations}
Substituting this in \eqref{dDS} results in
\begin{subequations}\label{redYW}
\begin{eqnarray}
   \label{redYWa}  U_t+ \sigma (US_y)_y+\frac{\dot g}{g}U&=&0,\\
   \label{redYWb}  S_t+\frac{\sigma}{2}S_y^2-\delta \Phi&=&-\frac{m^2}{2g^2},\\
   \label{redYWc} U_{yy}+\Phi_{yy}&=&\frac{g''}{\sigma\delta g} \, .
 \end{eqnarray}
\end{subequations}
From \eqref{redYWc} and \eqref{redYWb} we find
\begin{subequations}
\begin{eqnarray}
                   &&\Phi(y,t)=-U(y,t)+\frac{\ddot g(t)}{2\sigma \delta g(t)}y^2+f_1(t)y+f_0(t),\\
\label{formulaU}   &&U(y,t)=-\frac{1}{\delta}S_t-\frac{\sigma}{2\delta}S_y^2+\frac{\ddot g(t)}{2\sigma \delta g(t)}y^2+f_1(t)y+f_0(t)-\frac{m(t)^2}{2\delta g(t)^2} \, .
    \end{eqnarray}
\end{subequations}
Substituting \eqref{formulaU} into \eqref{redYWa}, it follows that $S(y,t)$ should satisfy the PDE
\begin{align}\label{red1}
\Big(\frac{ \ddot g}{2 g}y^2+\sigma\delta f_1y+\sigma\delta f_0-\frac{\sigma  m^2}{2g^2}\Big)S_{yy}-\sigma\Big(  S_t+\frac{3}{2}\sigma S_y^2\Big)S_{yy}-S_{tt}-2\sigma   S_y S_{yt}-\frac{\sigma \dot g}{2g}S_y^2    \nonumber\\
+\Big(\frac{ \ddot g}{ g}y+\sigma  \delta f_1\Big)S_y-\frac{\dot g}{g}S_t+\frac{\sigma \dddot g}{2 g}y^2+\delta(\dot f_1+f_1\frac{\dot g}{g})y+\delta(\dot f_0+f_0\frac{\dot g}{g})+\frac{m^2 \dot g}{2g^3}-\frac{m\dot m}{g^2}=0.
\end{align}
In order to find an exact solution of the highly nonlinear equation  \eqref{red1} we  introduce another equation
\begin{equation}\label{Syt}
S_t+\frac{3}{2}\sigma S_y^2=0,
\end{equation}
which is compatible with the given one. In other words, we look at the possibility of a nontrivial common  solution of \eqref{red1} and \eqref{Syt}.
Eq. \eqref{Syt} admits the separable solution
\begin{equation}\label{Sform}
S(y,t)=\frac{(y+\ell_1)^2}{6 \sigma  t+ \ell_2}
\end{equation}
where $\ell_1$, $\ell_2$ are integration constants. When we substitute \eqref{Sform} in \eqref{red1} we obtain a second degree polynomial in $y$ with coefficients in  $t$ as a compatibility condition, which has to vanish identically so that \eqref{red1} holds for \eqref{Sform}. The coefficient of $y^2$ is zero if $g(t)$ satisfies the equation
\begin{equation}
(6\sigma t+\ell_2)^3 \dddot g+6\sigma(6\sigma t+\ell_2)^2 \ddot g+8 (6\sigma t+\ell_2)\dot g-48\sigma  g=0.
\end{equation}
The general solution of this Euler type equation is
\begin{equation}
g(t)=K_1Q^{1/3}+K_2Q^{2/3}+K_3Q, \quad Q(t)=6\sigma t+\ell_2,
\end{equation}
where $K_1$, $K_2$, $K_3$ are arbitrary constants. The coefficients of $y^1$ and $y^0$ vanish when
\begin{align}
&f_0(t)=\frac{1}{2\delta}\frac{m^2}{g^2}+\frac{C_0}{g}Q^{-1/3}+\frac{C_1\ell_1}{g}Q^{-2/3}-\frac{4\ell_1^2\sigma}{\delta g}(K_1Q^{-5/3}+K_2Q^{-4/3}),\\
&f_1(t)=\frac{1}{g}\Big[C_1 Q^{-2/3}-\frac{8\ell_1\sigma}{\delta}(K_1Q^{-5/3}+K_2Q^{-4/3})\Big]
\end{align}
with the arbitrary constants $C_0$ and $C_1$.  Therefore we are able to find an exact solution of \eqref{dDS} depending on one arbitrary function $m(t)$. Explicitly, solution to \eqref{dDS} is
\begin{subequations}
\begin{eqnarray}
       u&=&U(y,t)=\frac{4\sigma (y+\ell_1)^2}{\delta (6\sigma t+\ell_2)^2}+\frac{\ddot g(t)}{2\sigma \delta g(t)}y^2+f_1(t)y+f_0(t)-\frac{m(t)^2}{2\delta g(t)^2},\\
       s&=&\frac{(y+\ell_1)^2}{6 \sigma  t+ \ell_2}+\frac{\dot g(t)}{2g(t)}x^2+\frac{m(t)}{g(t)}x, \\
    \phi&=&-U(y,t)+\frac{\ddot g(t)}{2\delta g(t)}x^2+\frac{\dot m(t)}{\delta g(t)}x+\frac{\ddot g(t)}{2\sigma \delta g(t)}y^2+f_1(t)y+f_0(t).
 \end{eqnarray}
\end{subequations}

We note that the symmetry algebra of $u_t=u_y^2$  is ten-dimensional.
One of its elements  is
$X_4=2t\gen y-y\gen u$ and generates a Galilei transformation.
The action of $X_4$ on a solution $u_0(y,t)$  gives the solution
\begin{equation}\label{sol2}
 u(y,t)=u_0(\tilde{y},\tilde{t})-\frac{\epsilon y^2}{4(1+\epsilon t)}, \quad \tilde{y}=\frac{y}{1-\epsilon t},  \quad \tilde{t}=\frac{t}{1-\epsilon t},
\end{equation}
and in particular on the separable solution
\begin{equation}
u_0=\frac{(y+c_2)^2}{c_1-4t}.
\end{equation}
gives the less trivial solution
\begin{equation}\label{sol3}
 u=\frac{(4-\epsilon c_1)y^2+8c_2 y+4c_2^2(1+\epsilon t)}{4[c_1+(\epsilon c_1-4)t]}.
\end{equation}
By scaling with respect to $t$ or $y$, we can write
\begin{equation}\label{sol4}
 S(y,t)=\frac{(4-\epsilon c_1)y^2+8c_2 y+2c_2^2(2-3\epsilon \sigma t)}{2[2c_1+ 3 \sigma(4-\epsilon c_1) t]}.
\end{equation}
as solution to \eqref{Syt}. As with \eqref{Sform}, this solution can be used to find other compatible solutions of \eqref{red1}.

\subsection{Reductions through two-dimensional subalgebras}
Now we  will perform symmetry reductions to ordinary differential equations (ODEs) using  the two-dimensional subalgebras $\{\mathscr{X}(t),\mathscr{Y}(g)+\mathscr{W}(m)\}$ and $\{\mathscr{X}(1),\mathscr{Y}(g)+\mathscr{W}(m)\}$. A two-dimensional subalgebra can be either  abelian  or non-abelian. We determine the functions $g(t)$  and $m(t)$ so that these two canonical forms are realized. Using the commutation relations for the generators given above in \eqref{XY} and \eqref{XW} we find the following four two-dimensional subalgebras, each depending on two arbitrary  parameters,
\begin{equation}\label{2-dim}
  \begin{split}
     & L_{2.1}=\{\mathscr{X}(t), a_1\mathscr{Y}(t^{3/2})+b_1 \mathscr{W}(t)\}, \quad
       L_{2.2}=\{\mathscr{X}(1), a_2\mathscr{Y}(e^t)+b_2 \mathscr{W}(e^t)\},   \\
     &L_{2.3}=\{\mathscr{X}(t), a_3\mathscr{Y}(t^{1/2})+b_3 \mathscr{W}(1)\},  \quad
     L_{2.4}=\{\mathscr{X}(1), a_4\mathscr{Y}(1)+b_4 \mathscr{W}(1)\},
  \end{split}
\end{equation}
with $L_{2.1}, L_{2.2}$ being non-abelian and $ L_{2.3}, L_{2.4}$  abelian.

\noindent \textbf{Reduction with $L_{2.1}$}:

By finding  invariants, we see that the ansatz
\begin{equation}\label{}
  \begin{aligned}
&u=\frac{1}{t}\, U(r),\\
&s=\frac{3x^2}{4t}+\frac{b_1 x}{a_1 \sqrt{t}}+S(r),  \\
&\phi = \frac{3x^2}{8\delta t^2}  + \frac{b_1 x}{a_1\delta t^{3/2}} + \frac{1}{t}\Phi(r)
\end{aligned}
\end{equation}
with $r=\displaystyle \frac{y}{\sqrt{t}}$ provides reduction to a system of ODEs
\begin{equation}\label{}
  \begin{aligned}
2\sigma(U S')'-rU'+U &=0,    \\
\frac{\sigma}{2} (S')^2 -\frac{r}{2}S' -\delta \Phi +\frac{b_1^2}{2a_1^2}  &=0,     \\
U''+\Phi''&=\frac{3\sigma}{4\delta} \, .
\end{aligned}
\end{equation}
Solving this system, we get
\begin{equation}\label{}
  \begin{aligned}
U(r)&=-\frac{\sigma}{2\delta}(S')^2+\frac{r}{2\delta}S'+\frac{3\sigma}{8\delta}r^2+c_1r+c_2-\frac{b_1^2}{2\delta a_1^2},\\
\Phi(r)&=\frac{\sigma}{2\delta}(S')^2-\frac{r}{2\delta}S'+\frac{b_1^2}{2\delta a_1^2},
\end{aligned}
\end{equation}
where $S(r)$ satisfies
\begin{align}\label{S-eq}
&\Big(\frac{\sigma}{4}r^2+2\delta c_1r+2c_2\delta-\frac{b_1^2}{a_1^2}\Big)\sigma S'' + 3\sigma rS'S''\nonumber \\
&-3(S')^2S''+\frac{\sigma}{2}(S')^2+(\frac{3r}{2}+2c_1\delta\sigma)S'-\frac{3\sigma}{8}r^2+c_2\delta-\frac{b_1^2}{2a_1^2}=0 .
\end{align}
Eq. \eqref{S-eq} equation admits a quadratic solution $S(r)=a r^2+br+c$. This occurs under the following conditions between the coefficients of the identity $k_2r^2+k_1r+k_0\equiv0$ required by \eqref{S-eq}:
\begin{subequations}
\begin{eqnarray}
       k_0&=&(6 a-\frac{\sigma}{2})b^2-c_2\delta(4a\sigma+1)-2bc_1\delta\sigma+\frac{(4a\sigma+1)b_1^2}{2a_1^2}=0,\\
       k_1&=&24 b a^2-8\sigma(b+c_1\delta)a-\frac{3 b}{2}=0,\\
       k_2&=&24  a^3-14\sigma a^2-\frac{7}{2}a+\frac{3\sigma}{8}=0 \, .
\end{eqnarray}
\end{subequations}
Solving this system we find
\begin{equation}
a=\{\frac{3\sigma}{4}, \frac{\sigma}{12}, -\frac{\sigma}{4}\},  \quad b=\frac{16\sigma a c_1\delta}{48a^2-16\sigma a-3}, \quad c_2=\frac{b_1^2}{2\delta a_1^2}+\frac{96ac_1^2 \delta}{(48a^2-16 \sigma a-3)^2}.
\end{equation}

Finally we find the two-parameter ($b_1,c_1$) solution of \eqref{dDS} in three distinct forms for each value of $a$ as
\begin{equation}\label{}
  \begin{aligned}
&u=\frac{1}{\delta }(a-2\sigma a^2+\frac{3\sigma}{8})\frac{y^2}{ t^2}+(c_1+\frac{b}{2\delta}-\frac{2ab\sigma}{\delta})\frac{y}{t^{3/2}}
  +(c_2-\frac{b_1^2}{2\delta a_1^2}-\frac{\sigma b^2}{2\delta})\frac{1}{t},\\
&s=\frac{3x^2}{4t}+\frac{b_1 x}{a_1 \sqrt{t}}+a\frac{y^2}{t}+b\frac{y}{\sqrt{t}}+c,  \\
&\phi = \frac{3x^2}{8\delta t^2}  + \frac{b_1 x}{a_1\delta t^{3/2}} +
\frac{a}{\delta}(2a\sigma-1)\frac{y^2}{t^2}+\frac{b}{2\delta}(4a\sigma-1)\frac{y}{t^{3/2}}+(\frac{b_1^2}{2a_1^2}+\frac{b^2\sigma}{2})\frac{1}{\delta t} \, .
\end{aligned}
\end{equation}

\noindent \textbf{Reduction with $L_{2.2}$}:

Invariance under $L_{2.2}$ leads to
\begin{equation}\label{}
  \begin{aligned}
&u=U(y),\\
&s=\frac{b_2}{a_2}x+\frac{x^2}{2}+S(y),  \\
&\phi = \frac{b_2}{\delta a_2} x + \frac{1}{2\delta} x^2 + \Phi(y)
\end{aligned}
\end{equation}
and to the reduced system
\begin{equation}\label{}
  \begin{aligned}
\sigma (US')' +U&=0,\\
\frac{\sigma}{2}(S')^2  -\delta \Phi +\frac{b_2^2}{2a_2^2} &=0,  \\
U''+\Phi''&=\frac{\sigma}{\delta} \, .
\end{aligned}
\end{equation}
Some work on these equations results in
\begin{equation}\label{}
  \begin{aligned}
&U(y)=-\frac{\sigma}{2\delta} (S')^2+\frac{\sigma}{2\delta}y^2+c_1y+c_2,\\
&\Phi(y)=\frac{\sigma}{2\delta} (S')^2+\frac{b_2^2}{2\delta a_2^2},
\end{aligned}
\end{equation}
where $S(y)$ satisfies the ODE
\begin{equation}
( y^2+2 c_1\sigma \delta y + 2c_2 \sigma \delta)S''-3 (S')^2 S''+2( y+c_1\sigma\delta)S'-\sigma (S')^2+\sigma y^2+2c_1\delta y+2c_2 \delta=0.
\end{equation}
Again,  a polynomial solution to this equation in the form $S(y)=ay^2+by+c$ is possible. This implies an expression of the form  $k_2y^2+k_1y+k_0\equiv 0$, where
\begin{subequations}
\begin{eqnarray}
       k_0&=& b^2 (\sigma+6a)-2c_2\delta(1+2\sigma a)-2\sigma bc_1\delta,\\
       k_1&=&24a^2b-2c_1\delta+4ab\sigma-8\sigma ac_1\delta-2b,\\
       k_2&=&(4a^2-1)(6a+\sigma) \, .
\end{eqnarray}
\end{subequations}
Solving $k_2=0$ we find  three possible values  $\displaystyle a=\{-\frac{\sigma}{6}, -\frac{1}{2},\frac{1}{2}\}$. When $(\sigma,a)\neq(1,-\frac{1}{2})$ and $(\sigma,a)\neq(-1,\frac{1}{2})$, the restrictions   $k_1=0$ and $k_0=0$ are solved to give
\begin{equation}
b=\frac{c_1\delta(1+4\sigma a)}{12a^2+2\sigma a-1}, \quad c_2=\frac{3\delta c_1^2 (4a+\sigma)}{2(12a^2+2\sigma a-1)^2},
\end{equation}
where $c$ and $c_1$ remain arbitrary. In case $(\sigma,a)=(1,-\frac{1}{2})$ and $(\sigma,a)=(-1,\frac{1}{2})$, we find $b=-c_1\delta$ and $c$, $c_1$, $c_2$ are arbitrary. We may express the solution in a unified form
\begin{equation}\label{}
  \begin{aligned}
&u=-\frac{\sigma}{2\delta} (2ay+b)^2+\frac{\sigma}{2\delta}y^2+c_1y+c_2,\\
&s=\frac{b_2}{a_2}x+\frac{x^2}{2}+ay^2+by+c,  \\
&\phi = \frac{b_2}{\delta a_2} x + \frac{1}{2\delta} x^2 + \frac{\sigma}{2\delta} (2ay+b)^2+\frac{b_2^2}{2\delta a_2^2} \, .
\end{aligned}
\end{equation}

\noindent \textbf{Reduction with $L_{2.3}$}:

Reduction formula
\begin{equation}\label{}
  \begin{aligned}
&u=\frac{1}{t}\, U(r),\\
&s=\frac{x^2}{4t}+\frac{b_3 x}{a_3 \sqrt{t}}+S(r),  \\
&\phi = -\frac{x^2}{8\delta t^2}  + \frac{1}{t}\Phi(r)
\end{aligned}
\end{equation}
with $r=\displaystyle \frac{y}{\sqrt{t}}$
leads to  the reduced system of ODEs
\begin{equation}\label{}
  \begin{aligned}
2\sigma(US')' -(rU)' &=0,    \\
\frac{\sigma}{2}(S')^2   -\frac{r}{2}S'  -\delta \Phi  +   \frac{b_3^2}{2a_3^2} &=0,     \\
U''+\Phi''&=-\frac{\sigma}{4\delta} \, .
\end{aligned}
\end{equation}
After some calculations, we see that $U(r)$ is a solution of the cubic  equation
\begin{equation}
4\delta U^3+\Big[\frac{2b_3^2}{a_3^2}-4\delta (c_1r+c_2)\Big]U^2+\frac{1}{2}\sigma c_3^2=0
\end{equation}
and
\begin{equation}\label{}
  \begin{aligned}
\Phi(r)&=-U(r)-\frac{\sigma}{8\delta}r^2+c_1r+c_2,\\
S(r)&=\frac{c_3}{2\sigma}\int \frac{dr}{U(r)}+\frac{r^2}{4\sigma}+c_4  .
\end{aligned}
\end{equation}
When  $c_1=0$, we find $\displaystyle U(r)=U_0$, a constant, which is a root of the cubic equation
\begin{equation}
4\delta U^3+\big(\frac{2b_3^2}{a_3^2}-4c_2\delta \big)U^2+\frac{1}{2}\sigma c_3^2=0 \, .
\end{equation}
Therefore, a solution to \eqref{dDS}  is obtained as
\begin{equation}\label{}
\begin{aligned}
u&=\frac{U_0}{t},\\
s&=\frac{x^2}{4t}+\frac{b_3x}{a_3\sqrt{t}}+\frac{y^2}{4\sigma t}+\frac{c_3 y}{2\sigma U_0 \sqrt{t}}+c_4,\\
\phi&=-\frac{1}{8\delta t^2}(x^2+\sigma y^2)+\frac{c_2-U_0}{t} \,.
\end{aligned}
\end{equation}

\noindent \textbf{Reduction with $L_{2.4}$}:

For this subalgebra we have
\begin{equation}
u=U(y),\quad s=\frac{b_4}{a_4} x+S(y),  \quad  \phi = \Phi(y)
\end{equation}
and the reduced system is
\begin{equation}\label{}
  \begin{aligned}
(US')'&=0,\\
 \frac{\sigma}{2}(S')^2  -\delta \Phi  + \frac{b_4^2}{2a_4^2}  &=0,  \\
U''+\Phi''&=0.
\end{aligned}
\end{equation}
We see that $U$ is a solution to the algebraic equation
\begin{equation}
2\delta U^3+[\frac{b_4^2}{a_4^2}-2\delta (c_1y+c_2)]U^2+\sigma c_3^2=0
\end{equation}
and
\begin{equation}\label{}
  \begin{aligned}
&\Phi(y)=-U(y)+c_1y+c_2,\\
&S(y)=c_3\int\frac{dy}{U(y)}+c_4.
\end{aligned}
\end{equation}

Finally, we comment that most of the solutions obtained in this Section are polynomials in $x$, and $y$, but with algebraic coefficients in $t$. Symmetry group transformation \eqref{sym-a-1} (or its subgroup \eqref{sym-gen}) can be applied to these solutions to construct other  solutions depending additionally on four arbitrary functions of $t$ and a new parameter.

%\section{Conclusion}

\section*{Acknowledgement}
We are grateful to the referee for drawing our attention  to  an issue leading to an incorrect identification of the symmetry algebra.

%\bibliography{dDS}
%\bibliographystyle{unsrt}
%\bibliographystyle{plain}

\end{document}